# Anomalous thermoelectric power of $Mg_{1-x}Al_xB_2$ system with x = 0.0 to 1.0

Monika Mudgel, V. P. S. Awana,* R. Lal, and H. Kishan
Superconductivity and Cryogenics Division, National Physical Laboratory, Dr. K.S. Krishnan Marg, New Delhi-110012, India

L. S. Sharth Chandra, V. Ganesan and A.V. Narlikar
UGC-DAE Consortium for Scientific Research, University Campus, Khandwa Road, Indore-452017, India

G. L. Bhalla
Department of Physics and Astrophysics, University of Delhi, Delhi-110007, India

**Abstract**

Thermoelectric power, S(T) of the $Mg_{1-x}Al_xB_2$ system has been measured for x = 0.0, 0.1, 0.2, 0.4, 0.6, 0.8 and 1.0. XRD, resistivity and magnetization measurements are also presented. It has been found that the thermoelectric power is positive for x ≤ 0.4 and is negative for x ≥ 0.6 over the entire temperature range studied up to 300 K. The thermoelectric power of x ≤ 0.4 samples vanishes discontinuously below a certain temperature, implying existence of superconductivity. In general, the magnitude of the thermoelectric power increases with temperature up to a certain temperature, and then it starts to decrease towards zero base line. In order to explain the observed behavior of the thermoelectric power, we have used a model in which both diffusion and phonon drag processes are combined by using a phenomenological interpolation between the low and high temperature behaviors of the thermoelectric power. The considered model provides an excellent fit to the observed data. It is further found that Al doping enhances the Debye temperature.

*PACS Nos. 74.25.fy, 74.62.Dh, 74.70.Ad*

Keywords: $MgB_2$ superconductor, Al substitution, Thermoelectric power, Debye Temperature, Phonon contribution

*: Corresponding author for further queries: Dr. V.P.S. Awana
Fax No. 0091-11-25726938: Telephone. 0091-11-25748709
e-mail: awana@mail.nplindia.ernet.in
Web page: www.freewebs.com/vpsawana/

## I. INTRODUCTION

With the advent of high temperature superconductivity in 1986 [1] the possibility of electron–phonon interaction as a superconductivity mechanism became scant. However, the situation changed after the discovery of superconductivity at 39 K in $MgB_2$ [2], where it was realized that mechanism of superconductivity could yet be based on the electron-phonon interaction within the strong coupling limits [3]. This was perceived from the fact that both Mg and B being light elements their lattice contributions could be sufficiently strong to promote the relatively high $T_c$ observed [4]. Besides, the lattice of $MgB_2$ is stretched in *c*-direction in comparison to other same structure borides viz., $TaB_2$, $AlB_2$, $ZrB_2$ or $MoB_2$ [5-7]. Stretched lattice may result in its instability and hence further more contribution to phonon interaction. It seems, the basic stretched lattice structure of $MgB_2$ being constructed from relatively lighter elements Mg and B is responsible for strong electron-phonon interaction. $MgB_2$ possess simple hexagonal $AlB_2$-type structure with space group $P_6/mmm$. It contains the graphite-type boron layers, which are separated by hexagonal close-packed layers of magnesium. The magnesium atoms are located at the center of hexagons formed by boron. The spacing between the boron planes is significantly larger than the in-plane B-B spacing. In fact, the characteristic *c/a ratio is* ~ 1.14 in $MgB_2$, while it is ~ 1.08 in $AlB_2$.

Another important difference between $MgB_2$ and $AlB_2$ is that in $MgB_2$ the cation ($Mg^{2+}$) is divalent while in $AlB_2$ the cation ($Al^{3+}$) is trivalent. This means that if we gradually replace Mg by Al, the population of the holes in the two-dimensional (2D) $\sigma$-band will start decreasing. In fact, it has been shown by band structure calculations that the $\sigma$-band is placed lower than the Fermi energy in $AlB_2$ [3,8]. So substitution of Mg by Al will fill the $\sigma$-band completely (with electrons) even before a 100% substitution level. In this situation, only $\pi$-band will take part in various physical processes so that the system will no more be a two-band ($\sigma$-band and $\pi$-band) system. This two-band to one-band system crossover by increasing substitution of Mg by Al in $MgB_2$ prompted us to carry out the present study of $Mg_{1-x}Al_xB_2$ system for $x = 0.0$ to $x = 1.0$. Earlier reports on $Mg_{1-x}Al_xB_2$ were limited to $x \leq 0.4$ [5], $x \leq 0.1$ [9] and $x \leq 0.5$ [10] only. Slusky et al [5] have studied crystal structure and magnetization of $Mg_{1-x}Al_xB_2$ for $x \leq 0.4$. Their main finding is that Al doping near 10% causes partial collapse of the spacing between boron layers and that the superconductivity has a close connection with such a structural instability. Lorenz et al [9] have measured the thermoelectric power (TEP) of the Al-substituted samples up to x = 0.1. They made an effort to explain the observed behavior of TEP on the basis of the *parabolic one-band model*. Very recently, Monika et al [10] have presented a detailed X-ray diffraction study along with the resistivity and magnetization measurements for the $Mg_{1-x}Al_xB_2$ for $x \leq 0.5$. Mainly the existence of superstructure is shown in the $Mg_{0.5}Al_{0.5}B_2$ system.

In this paper we focus on the thermoelectric power of $Mg_{1-x}Al_xB_2$ for Al concentrations ranging from x = 0.0 to x = 1.0. This wide range of x values allow us to study the behavior of $Mg_{1-x}Al_xB_2$ system for varying contributions of the σ - band. Recently Souma and Takahashi [11] (cf. Figs. 1 & 4 of these authors) had shown the π-band structure to be essentially same for both $MgB_2$ and $AlB_2$. So, we can consider the contributions of the π-band to be similar for all x. As mentioned above, TEP of $Mg_{1-x}Al_xB_2$ has been observed earlier also [9] but for $x \leq 0.1$. We shall not only present TEP data for $Mg_{1-x}Al_xB_2$ for a much wider range of x ($0.0 \leq x \leq 1.0$) but we shall also provide a reasonable explanation of the TEP behavior. In fact the explanation of the TEP behavior given by Lorenz et al [9] for the $Mg_{1-x}Al_xB_2$ system and also by Gahtori et al for the $Mg_{1-x}Fe_xB_2$ system [12] seems defective because of the following reasons. (1) The linear theoretical fits based on the parabolic one band model used by these authors do not pass through the origin of the temperature-TEP plane. This amounts to inconsistency because according to Eq. (3) of Lorenz et al and Eq. (12) of Gahtori et al [12], TEP should vanish for zero temperature. On the other hand, Lorenz et al find S = -1.0μV/K for T=0, and Gahtori et al find S = -1.2μV/K for T=0 for their respective $MgB_2$ samples. (Here S denotes TEP and T denotes temperature). These values are quite significant as the TEP of $MgB_2$ is of the order of 1.0 μV/K (near 50 K) for both the cases. (2) The agreement between the theoretical model and experimental results tends to become poorer with increasing temperature within the one-band model employed by Lorenz et al and Gahtori et al (cf. Fig. 2 of Ref. [9] and Fig. 3 of Ref. [12]). For a realistic explanation of the TEP, we, in the present paper, shall consider the phonon drag contribution also. Apart from this, we shall also consider the high-T behavior of the TEP simultaneously.

## II. EXPERIMENTAL DETAILS

Polycrystalline $Mg_{1-x}Al_xB_2$ samples with x = 0.0, 0.10, 0.20, 0.40, 0.60, 0.80 and 1.0 are synthesized by solid-state reaction route with ingredients of Mg, B, and Al. The Mg powder used is from *Reidel-de-Haen* of assay 99%. B powder is amorphous and *Fluka* make of assay 95-97%. The Al powder is from *Reidel-de-Haen* with above 93% assay. For synthesis of $Mg_{1-x}Al_xB_2$ samples, the nominal weighed samples are ground thoroughly, palletized, encapsulated in soft iron tube and put in a programmable furnace under flow of Argon at one atmosphere pressure. The temperature of furnace is programmed to reach $850^0$ C over 2 hours, hold at same temperature for two and a half hours, and subsequently cooled to room temperature over a span of 6 hours in the same Argon atmosphere. X- ray diffraction patterns were taken using Ni Filtered $CuK_\alpha$ radiation. Resistivity measurements were carried out by conventional 4-probe method. Thermoelectric power measurements were carried out by dc differential technique over a temperature range of 5 – 300 K, using a home made set up. Temperature gradient of ~1 K was maintained throughout the TEP measurements. Magnetization measurements are carried out with a *Quantum-Design* SQUID magnetometer MPMS-7.

## III. RESULTS AND DISCUSSION

Room temperature X-ray diffraction (XRD) patterns for the $Mg_{1-x}Al_xB_2$ with x = 0.0 to 1.0 are shown in Fig. 1. All the samples crystallize in simple hexagonal $AlB_2$-type structure with space group $P_6/mmm$. For pristine sample all characteristic peaks are indexed which are in well corroborated by literature [2,4,5,10]. With successive substitution of Al at Mg site in $Mg_{1-x}Al_xB_2$, till x = 0.40, though the structure (hexagonal) and space group $P_6/mmm$ remain the same, all the XRD peak positions are shifted towards higher angle side, indicating a decrease in lattice parameters. The upper and lower insets of Fig. 1 show shifts of (002) and (100) peak confirming the decrease of *a* and *c* parameter. For $MgB_2$; *a*=3.0857Å & *c*= 3.5230 Å, while for $AlB_2$; *a* = 3.0036 Å & *c* = 3.2519 Å. For other samples of the series both *a* and *c* lattice parameters have intermediate values according to Al content in them. The structural anomaly related with broadening of (002) peak up to x = 0.50 is described in detail by us elsewhere [10]. Here we have analyzed the samples up to full Al substitution. Beyond x = 0.50, some additional phases arise, as shown by * in Fig. 1, which might be due to $AlB_4$.

Resistivity versus temperature plots of the $Mg_{1-x}Al_xB_2$ series with x = 0.0, 0.10, 0.20, 0.40, 0.60, 0.80 & 1.0 are shown in Fig. 2. In the normal state i.e. above $T_c$, all samples show metallic behavior. The normal state ρ-T plot of our pure $MgB_2$ corresponds to $\rho(300)/\rho(T_c) \sim 3$, which agrees with that for the $MgB_2$ samples of Lorenz et al [9] and Gahtori et al [12]. Critical temperature $T_c$ (ρ = 0) for pristine sample is about 38 K. As we substitute Al, loss of superconductivity is observed in terms of decreasing critical temperatures [9,13,14]. In fact $T_c$ is 34, 30 & 7 K for Al content of 10%, 20% and 40% respectively. This means that, there is slow decrease in $T_c$ up to x = 0.20, which is followed by relatively much sharper decrement till x = 0.40. The samples are no longer superconducting (ρ =0) beyond x = 0.40, while sample with x = 0.60 exhibits a $T_c^{onset}$ only. The superconducting transition width is small up to x = 0.20 samples, but for x = 0.40 sample, $T_c^{onset}$ is 26.2 K, while $T_c$ (ρ = 0) is only 6.9 K. No superconductivity (in terms of ρ =0) is observed in $Mg_{0.5}Al_{0.5}B_2$ sample & beyond that. magnetization measurements (χ - T) for $Mg_{1-x}Al_xB_2$ series with x = 0.0, 0.04, 0.10, 0.20, 0.25 and 0.40 are shown in the inset of Fig. 2. The critical temperatures obtained from here ($T_c^{dia}$) are in close agreement with the critical temperature ($T_c^{\rho = 0}$) obtained from resistivity measurements except in the case of x = 0.40 for which a broad transition is seen and the saturation of moment is not seen down to 5 K. Qualitatively both ρ - T and χ - T measurements confirm a similar decrease in $T_c$ with Al doping. The decrease of $T_c$ with Al can easily be explained in terms of electron doping. Each Al atom provides an extra electron and results in the hole band filling. Electron doping raises the Fermi level to higher energies and hence density of states decreases at Fermi Level, which results in loss of superconductivity. It is believed that superconductivity in $Mg_{1-x}Al_xB_2$ is due to electron-phonon interaction, and that the $T_c$ suppression due to increasing Al is caused by the lowering of the σ-band [8]. Moreover, Lattice parameters also decrease continuously with Al doping. The c/a value for $MgB_2$ is 1.14 while the same is just 1.08 for $AlB_2$. So the $AlB_2$ lattice is quite compressed in comparison to $MgB_2$ This lattice strain also affects the electron phonon interactions and hence may be a secondary reason for the suppression of superconductivity.

The measured thermoelectric power of the $Mg_{1-x}Al_xB_2$ samples for x = 0.0, 0.1, 0.2, 0.4, 0.6, 0.8 and 1.0 are presented in Fig. 3 by using different symbols for different samples. For low x, i.e., x < 0.4, the behavior of the TEP is similar to that of the $Mg_{1-x}Al_xB_2$ samples of Lorenz et al [9], and to that of the $Mg_{1-x}Fe_xB_2$ samples of Gahtori et al [12]. In particular, there is signature of superconductivity in the behavior of TEP for the $x \leq 0.4$ samples. This is consistent with the magnetization and resistivity measurements shown in Fig. 2. The TEP of the $x \geq 0.6$ samples is negative for all the considered temperature values, signifying the ineffectiveness of the hole based σ-band. Also the TEP of the $x \geq 0.6$ samples does not show any feature of superconductivity.

In order to explain the observed behavior of TEP of the present $Mg_{1-x}Al_xB_2$ samples, we notice that in general this system involve two bands – the σ-band and the π-band. It is believed that the σ-band is primarily responsible for the occurrence of the superconductivity [8]. Then, since the observed behavior of TEP involves the effect of superconductivity (for $x \leq 0.4$), we may say that σ-band holes provide essential contribution to the TEP values in this doping range. Since, for higher x, the extra electrons donated by Al tend to make the σ-band ineffective beyond a certain doping level of Al, hence it may be argued that the TEP of the x > 0.4 samples will arise due to the π-band only. The features of the π-band are reported similar in $MgB_2$ and $AlB_2$ [11], this we will use below for setting a reasonable expression of the TEP.

Theoretically, two processes contribute to the TEP. The first is the diffusion process, and second is the phonon drag process. The contribution to TEP due to diffusion process is given by the Mott formula [15]

$$S_d = (\pi^2 k_B^2 T/3e) \left| [\partial \ln\sigma(\omega)/\partial\omega] \right| E_F \tag{1}$$

Here $k_B$ is the Boltzmann constant, e is carrier charge, σ (ω) is conductivity corresponding to the electron energy ω, and $E_F$ is Fermi Energy. In the low temperature limit, the carrier relaxation time is limited by impurity scattering. From the extrapolation of the resistivity of $Mg_{1-x}Al_xB_2$ (shown in Fig.2) to zero temperature, it turns out that the present samples correspond to significant resistivity at zero temperature. This means that there is an essential presence of impurity scattering in the present samples. Thus for low temperatures, where phonons have not yet started to play a significant role, we can use the conductivity formula [16]

$$\sigma = ne^2\tau/m \qquad (2)$$

Here n is carrier density, $\tau$ is carrier relaxation time and $m$ is carrier mass. For a three dimensional system, it is well known that Eqs. (1) and (2) lead to Eq. (3) below [15]

$$S_d = \pi^2 k_B^2 T/3eE_F \qquad (3)$$

In order to obtain an expression for diffusion thermoelectric power corresponding to a two-dimensional band, we follow the same steps as used in arriving at Eq. (3) from Eqs. (1) and (2). We use the 2D form of $\sigma$ from Eq. (2c) of Fukuyama [16] and find that formally Eq. (3) is valid for the 2D band also. We emphasize that this is only for parabolic 2D and 3D bands. The result may be different for other band structures. Since the TEP is additive (cf. Eq H8 of Bailyn [17]), the combined contribution of the 3D $\pi$-band and 2D $\sigma$-band of the $Mg_{1-x}Al_xB_2$ to TEP may be written as

$$S_d = (\pi^2 k_B^2 T/3|e|)(W_\sigma^{-1} - W_\pi^{-1}) = AT \qquad (4)$$

Here $W_\sigma$ is the separation of the Fermi level from the bottom of the hole-like $\sigma$-band, and $W_\pi$ is the separation of the Fermi level from the bottom of the $\pi$-band. A is a constant, independent of temperature. From the band structure calculations of An and Pickett [8], one finds that $W_\sigma$ =0.72 eV, and $W_\pi$ =7.4 eV. This means that the contribution of the $\pi$–band electrons to $S_d$ is about 10 times smaller than that of the $\sigma$–band holes.

As mentioned in introduction, the diffusion contribution to TEP leads to inconsistent explanation of the reported data [9,12]. In fact, we obtain TEP approximately equal to –1.0μV/K for T = 0, while it should have vanished according to Eq. (4) used in Refs. [9] and [12]. Since Eq. (4) is justified for $T\rightarrow 0$, and also we get inconsistency by the use of Eq. (4), it becomes necessary to consider the phonon drag contribution along with the diffusion term to the thermoelectric power. For the 3D $\pi$-band the phonon drag contribution will vary like $T^3$ for low T [18], and like $T^{-1}$ for high T [19]. Though for the intermediate temperature values, the behavior of the phonon drag TEP is given by a very complicated expression [17, 19], but still the same varies smoothly from low T values to the high T values. We thus hope that a simple interpolation will provide a reasonable phenomenology of the variation of the phonon drag contribution to TEP from $T\rightarrow 0$ to $T\rightarrow\infty$. Since TEP varies as $T^3$ for $T\rightarrow 0$ and as $T^{-1}$ for $T\rightarrow\infty$, we consider the interpolation

$$S_{pd,\,\pi} = T^3/(B+CT^4) \qquad (5)$$

for the 3D π- band. Here the suffix 'pd' on S implies phonon-drag and the suffix 'π' implies the 3D π- band.

From Eq. (5), we see that, $S_{pd,\pi}\rightarrow T^3/B$ for $T\rightarrow 0$. Here B should vary like $\theta_D^3$ [18]. Here $\theta_D$ is the Debye temperature. C is also a constant, but not dependent on $\theta_D$.

In order to work out an expression for the phonon drag contribution of the 2D σ-band, we notice that the 2D character of the σ-band will allow the phonons to drag only in a 2D plane. As a result of this, the reason that led to a $T^3$ variation of the phonon drag contribution from a 3D band, will lead to a $T^2$ variation for a 2D band. Thus using the interpolation similar to that of Eq. (5), we may express the phonon drag TEP due to the σ-band by

$$S_{pd,\,\sigma} = T^2/(D+ET^3) \qquad (6)$$

Notice that here D is expected to vary like vary like $\theta_D^2$. However, since the σ-band changes with Al doping in $MgB_2$, D may also have a dependence on the electronic structure. We shall see below what the present data inform in this connection.

Combining all the contributions from Eqs. (4) - (6), we obtain finally the following expression for the TEP of the $Mg_{1-x}Al_xB_2$ samples.

$$S = AT + T^3/(B+CT^4) + T^2/(D+ET^3) \qquad (7)$$

We emphasize that we have not included high-T contribution in Eq. (7) due to the diffusion process. The reason for this is that for higher values of temperature, the carrier conductivity will depend upon the electron phonon interaction [12]. In fact, the conductivity due to electron-phonon interaction is a complicated expression [12]. We have to take logarithm of this expression (cf. Eq. (1)), and then a differentiation. This is almost an impractical task. So here we have not considered the effect of the electron-phonon interaction on the diffusion of the carriers.

We now turn to the explanation of the observed TEP data. Using Eq. (7) we have fitted the observed data. The fitting parameters are presented in Table 1. The parameter E acquires practically zero value. From Fig. 3 we see that Eq. (7) provides an excellent agreement of the theoretical Eq. (7) with the observed data in the whole temperature range except for the x = 0.8 and 1.0 samples. In fact, the x = 0.8 and 1.0 samples also show good agreement qualitatively. In view of such an agreement we would like to know about the relative contributions of the diffusion process and phonon drag process. For this purpose, we take a specific temperature T = 100 K, and consider values of S for the x = 0 and x = 0.6 samples. From the calculated values, it turns out that the x = 0 sample corresponds to $S_d$ = 4.0 μV/K, $S_{pd,\,\sigma}$ = -0.46 μV/K and $S_{pd,\,\pi}$ = -1.09 μV/K. We see that the phonon drag corresponds to significant contribution. Here we would like to clarify that depending upon the electronic structure; the contribution of the phonon drag process to TEP may be positive or negative irrespective of the charge of the carriers. This argument is based on the work of Bailyn [20].

We now consider the relative contribution of the diffusion process and phonon drag process to TEP for the x = 0.06 sample at T=100 K. We find $S_d$ = -0.17 μV/K, $S_{pd,\,\sigma}$ = 0 μV/K and $S_{pd,\,\pi}$ = -0.29 μV/K. From these values it is clear that the phonon drag process is a dominating process for the TEP contribution in the x = 0.6 sample. In fact, as is clear from the values of A of Table 1, the diffusion contribution is almost negligible for the x ≥ 0.6 samples. The main reason for this is the very large values of $W_\pi$ (see above).

In order to extract more information from the parameters of Table 1, we first of all notice that the values of the parameter A are significant only for the superconducting (x ≤ 0.4) samples. Moreover, the value of A decreases with increasing x. This is in contradiction with that of Lorenz et al [9] who find an increasing linear- T slope of TEP. An obvious reason for this is the modification of the theoretical TEP due to phonon drag. In fact, Lorenz et al have not considered phonon drag. In order to see how phonon drag may affect the value of the linear-T co-efficient in Eq. (7), we start from the fact that the phonon drag contribution to TEP varies from a $T^3$-like behavior for low-T to $T^{-1}$ like behavior for high-T. Since the low-T to high-T variation of the phonon drag contribution takes place smoothly, we will come across $T^2$-like, T-like and $T^0$ like variations while going from low T-side to high T-side. In fact the region near the peak of S may be considered to be a $T^0$ like (constant) behavior while below that it is a T-like behavior. The T-like portion will get combined with the T-like diffusion contribution. Thus the value of A will be modified than what it would have been in the absence of the phonon drag. For $MgB_2$ we find A = 0.04 $\mu V/K^2$, while Lorenz et al find a slope equal to 0.042 $\mu V/K^2$. These slopes are almost equal. In fact, using $W_\pi$ = 7.4 eV [8] in Eq. (4), we find that $W_\sigma \approx 0.57$ eV, which is the same as that obtained by Lorenz et al [9]. However, since the phonon drag affects the value of A, it is doubtful to treat the value of $W_\sigma$ as the value of the ($\sigma$-band) Fermi energy. We thus argue that the first term of Eq. (7) provides only the linear-T contribution to TEP where the diffusion process is modified by the phonon drag process.

We next consider the parameter B. As mentioned above (the magnitude of) B should vary like $\theta_D^3$. Let $\theta_{D0}$ be the Debye temperature of the $MgB_2$ sample. Then using the values of B from Table 1, we can estimate the Debye temperature $\theta_D$ (x) for different x with respect to $\theta_{D0}$. The values of $\theta_D$ (x)/ $\theta_{D0}$ obtained in this way are given in the last column of table I. From these values we see that $\theta_D$ (x) increases with Al content up to x = 0.4. Then $\theta_D$(x) start to decrease with further increase in x so that for the x = 0.8 sample $\theta_D$(x) takes the lowest value of 0.94 $\theta_{D0}$. For the $AlB_2$ sample, the Debye temperature is 55% higher than that of the $MgB_2$ sample. So, in general Al enhances the Debye temperature of the $Mg_{1-x}Al_xB_2$ system, although in a non-monotonic way. The Debye temperature depends mainly upon two factors. The first is the interatomic coupling, and second is the mass of the constituent atoms. While $\theta_D$(x) increases with the interatomic coupling, it decreases with the atomic mass. Since Al is heavier than Mg, we expect a decrease in $\theta_D$(x) due to the mass effect of Al . But table I shows that $\theta_D$(x) increases with Al content. So we may say that doping of Al enhances the interatomic coupling in $Mg_{1-x}Al_xB_2$. As the lattice parameters *a* and *c* are found to be decrease with Al, the atoms are coming closer with Al doping. This may be a possible reason for the enhancement of interatomic coupling in the considered samples. It is surprising that the Debye temperature of the x = 0.8 sample is lower than $\theta_{D0.}$ From Fig.2 we see that the resistivity of this sample has much higher slope values at different T than the other Al doped samples. This means that the situation in x= 0.8 sample is different, which is reflected in the observed TEP also.

As mentioned above, the parameter D also depends on $\theta_D$, like $\theta_D^2$. However, D is expected to involve the effect of varying features of the σ-band also with Al doping. In fact, if D were to depend on $\theta_D$ only, the factor $p = |D|\theta_{D0}^2/\theta_D^2$ should have been independent of x. Let us see what actually happens. From Table 1 we find that p = 2.16, 1.73, 1.33 and 1.53 in units of $10^4$ $K^3/\mu V$ for x = 0.0, 0.1, 0.2 and 0.4 respectively. Since p is not constant but varies significantly with x, we may say that there is a variation in the σ-band structure due to Al doping, and that the variation has affected the TEP behavior for different Al content.

## IV. Conclusions

In this paper, we have presented measurements of the thermoelectric power of the $Mg_{1-x}Al_xB_2$ system for different values of x ranging from zero to one. Measurements of XRD, magnetization and resistivity are also presented. The thermoelectric power of the $x \leq 0.4$ samples vanishes discontinuously below a certain temperature, implying existence of superconductivity. The thermoelectric power of the $x \geq 0.6$ samples is negative, and does not vanish below a finite temperature. This is argued to show that thermoelectric power does not indicate superconducting effect in the $x \geq 0.6$ samples. Another important feature of the thermoelectric power is that its magnitude, |S| starts to increase with temperature, and continues so up to a certain temperature. The temperature at which |S| is maximum, decreases in general with x, going to as low as $T \approx 80$ K, for the x = 0.8 sample.

In order to explain the observed behavior of the thermoelectric power, we have used a two-band model wherein one band is the 3D π-band and the other one is the 2D σ-band. We have considered both the diffusion process and the phonon drag process to arrive at an interpolation formula from the low-T behavior of the diffusion and phonon drag process and the high-T behavior of the phonon drag process. The interpolation formula provides an excellent agreement especially for the $x \leq 0.6$ samples. For the x= 0.8 and 1.0 sample, there is some quantitative disagreement but the qualitative behavior remains matching very well. We have found that the slope of S with respect to T is almost equal for the x = 0 sample to that of the Lorenz et al [9] sample.

Another important information from the present study is that in general the Al doping enhances the Debye temperature. The thermoelectric power of $AlB_2$ sample does not have a sizable straight-line portion towards low temperature, implying practically no diffusion contribution.

**Acknowledgement**

The authors from NPL would like to thank Dr. Vikram Kumar (Director, NPL) for showing his keen interest in the present work. One of us (AVN) thanks the Indian National Science Academy (INSA), New Delhi for Senior Scientist position. Two of us (MM and LSSC) would also thank CSIR for financial support in the form of (by providing JRF*)* fellowship.

**Figure Captions**

Fig.1 XRD patterns for $Mg_{1-x}Al_xB_2$ series (x = 0.0 to x = 1.0) Upper and lower insets show shifts of the (002) & (100) peaks respectively.

Fig. 2 Resistivity vs temperature plots for all $Mg_{1-x}Al_xB_2$ samples (x = 0.0 to x = 1.0), The inset shows the zero field cooled (ZFC) magnetization as a function of temperature for superconducting $Mg_{1-x}Al_xB_2$ (x = 0.0 to x= 0.40) samples.

Fig.3 Thermopower vs temperature plots in the temperature range 0 to 300 K for all samples of series $Mg_{1-x}Al_xB_2$ (x = 0.0 to 1.0). The experimental data points are shown by different symbols, while the theoretical fits to Eq. (7) are shown by solid lines.

Table1: Values of the parameters A, B, C and D of Eq. (7) for various values of Al content in $Mg_{1-x}Al_xB_2$. The values of E turn out to be less than $10^{-40}$ so it has been taken zero for all x in $Mg_{1-x}Al_xB_2$. The relative values of the Debye temperature $\theta_D/\theta_{Do}$ are also given. $\theta_D$ is the Debye temperature of the x = 0 sample.

| x | A (μV/K$^2$) | B (K$^4$/μV) | C (1/μV) | D (K$^3$/μV) | $\theta_D(x)/\theta_{D0}$ |
| --- | --- | --- | --- | --- | --- |
| 0.0 | 0.040 | $-5.91\times10^4$ | $-8.59\times10^{-3}$ | $-2.16\times10^4$ | 1.00 |
| 0.1 | 0.036 | $-6.84\times10^4$ | $-17.18\times10^{-3}$ | $-1.82\times10^4$ | 1.05 |
| 0.2 | 0.036 | $-10.29\times10^4$ | $-17.44\times10^{-3}$ | $-1.59\times10^4$ | 1.20 |
| 0.4 | 0.012 | $-37.56\times10^4$ | $-38.37\times10^{-3}$ | $-2.83\times10^4$ | 1.85 |
| 0.6 | -0.002 | $-27.97\times10^4$ | $-31.11\times10^{-3}$ | - | 1.68 |
| 0.8 | -0.001 | $-4.89\times10^4$ | $-6.43\times10^{-3}$ | - | 0.94 |
| 1.0 | 0.0 | $-21.83\times10^4$ | $-6.50\times10^{-3}$ | - | 1.55 |

Fig. 1

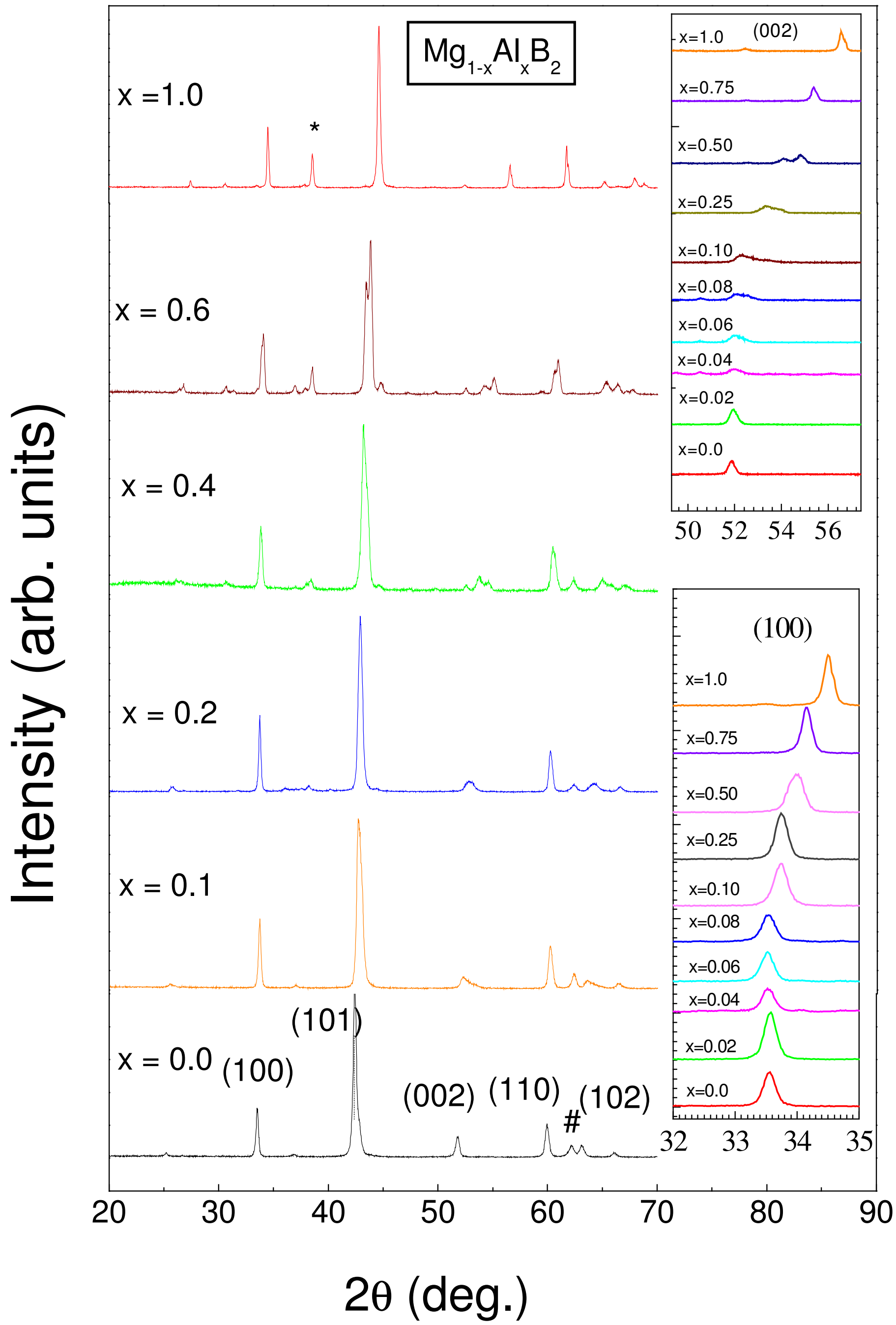

Fig. 2

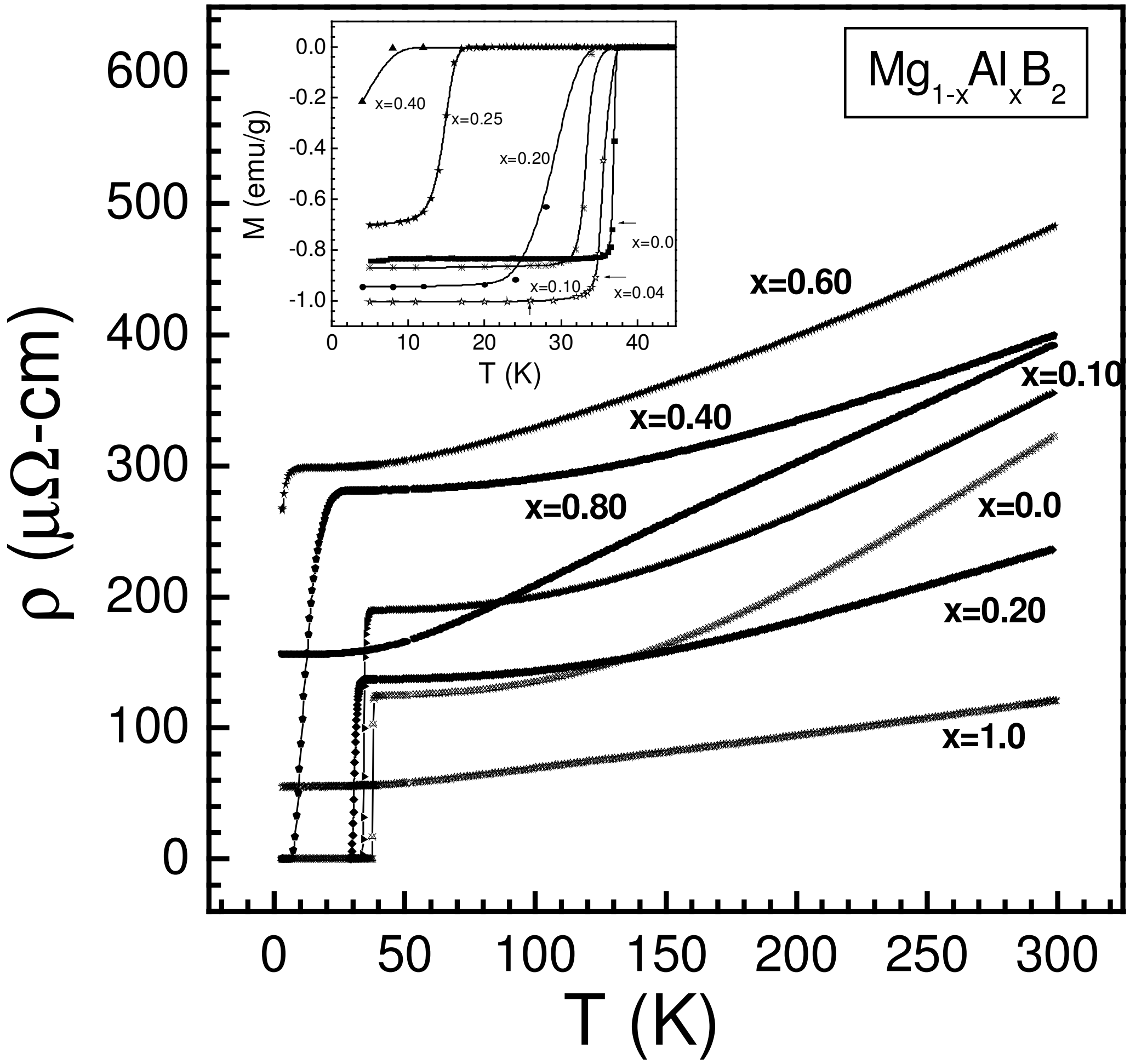

Fig. 3

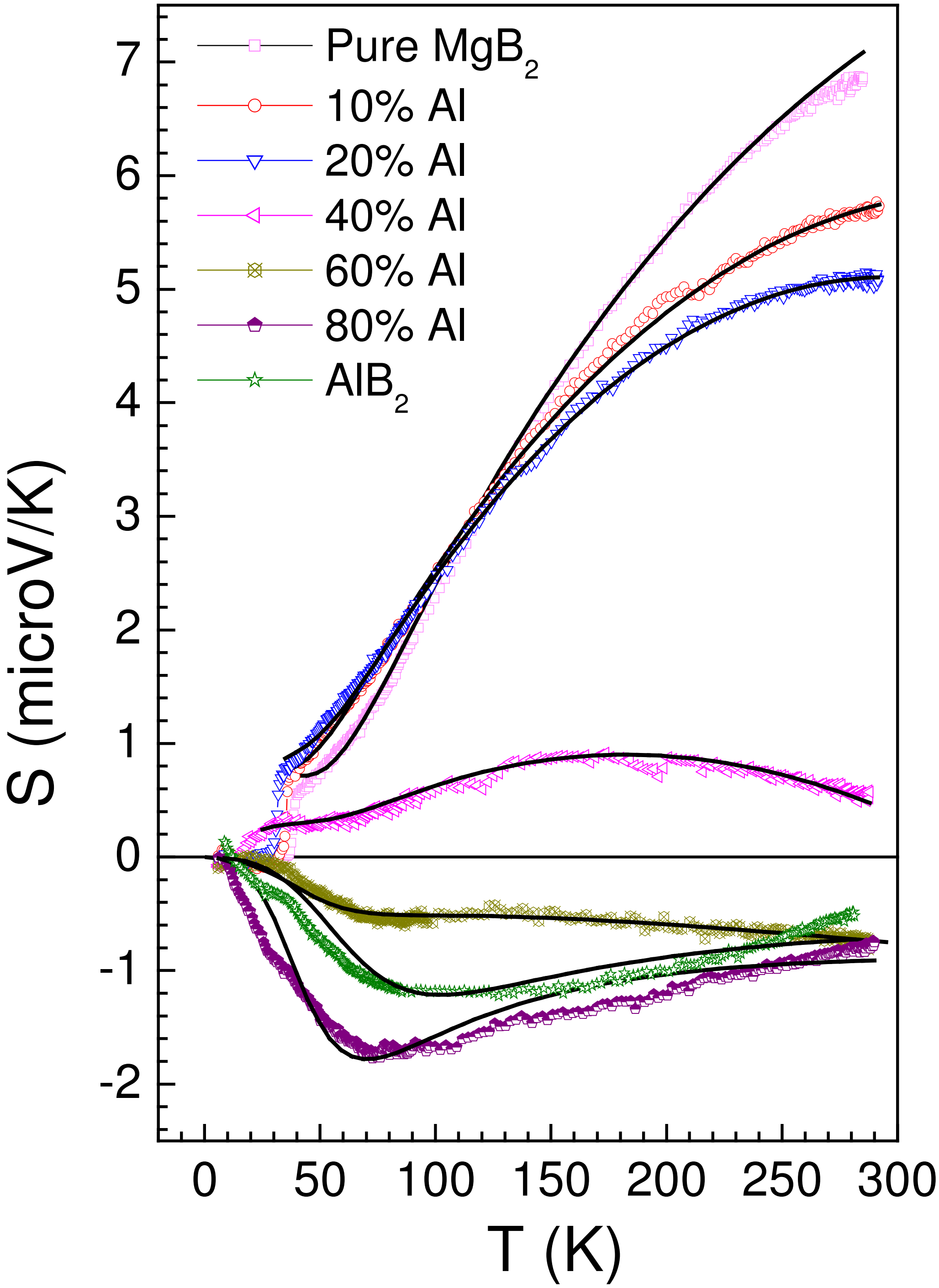